\documentclass[twocolumn,superscriptaddress,showpacs,aps,amsmath,amssymb,pra]{revtex4}

\newif\ifFigureFolder
\newif\ifArXivCompatible

\FigureFolderfalse			
\ArXivCompatibletrue		

\usepackage{subfigure}
\usepackage{graphicx}
\usepackage{color}

\ifArXivCompatible
\else 
	\usepackage{todonotes}
	\usepackage{hyperref}
\fi
\ifFigureFolder
	\graphicspath{{./figures/}}
\fi

\begin{document}

\title{Superconducting quantum metamaterials as active lasing medium: Effects of disorder}

\author{Martin Koppenh\"ofer}
\affiliation{Institut f\"ur Theoretische Festk\"orperphysik, Karlsruhe Institute of Technology, D-76131 Karlsruhe, Germany}

\author{Michael Marthaler}
\affiliation{Institut f\"ur Theoretische Festk\"orperphysik, Karlsruhe Institute of Technology, D-76131 Karlsruhe, Germany}

\author{Gerd Sch\"on}
\affiliation{Institut f\"ur Theoretische Festk\"orperphysik, Karlsruhe Institute of Technology, D-76131 Karlsruhe, Germany}
\affiliation{Institute of Nanotechnology, 
Karlsruhe Institute of Technology, D-76344 Eggenstein-Leopoldshafen, Germany}


\date{\today}

\newcommand{\ket}[1]{\left\vert #1 \right\rangle}
\newcommand{\bra}[1]{\left\langle #1 \right\vert}
\newcommand{\abs}[1]{\left\vert #1 \right\vert}
\newcommand{\erw}[1]{\left\langle #1 \right\rangle}
\newcommand{\erfc}{\operatorname{erfc}}
\renewcommand{\d}{\mathrm{d}}

\begin{abstract}
A metamaterial formed by superconducting circuits or quantum dots can serve as active lasing medium when coupled to a microwave resonator. For these artificial atoms, in contrast to real atoms, variations in their parameters cannot be avoided. In this paper, we examine the influence of disorder on such a multi-atom lasing setup. We find that the lasing process evolves into a self-organized stationary state that is quite robust against disorder. The  reason is that photons created by those atoms which are in or close to resonance with the resonator stimulate the emission also of more detuned atoms. Not only the number of photons grows with the number of atoms, but also the width of the resonance as function of the detuning. Similar properties are found for other types of disorder such as variations in the individual coupling. We present relations how the allowed disorder scales with the number of atoms and confirm it by a numerical analysis. We also provide estimates for the sample-to-sample variations to be expected for setups with moderate numbers of atoms.
\end{abstract}

\maketitle

\section{Introduction}
Lasers are the standard sources of coherent light 
with a wide range of applications \cite{LambScully-LaserPhysics}. Their basic components are a resonator that stores photons and selects particular modes, an optically active medium that emits photons coherently into the resonator by stimulated emission, and a pumping process that establishes a population inversion in the medium
\cite{ScullyZubairy-QuantumOptics}. A large variety of systems can serve as optically active medium. This includes natural atoms or semiconductor devices 
\cite{ThompsonSemiconductorLaser}, but further systems have been proposed and studied experimentally. These include strongly coupled single or few Josephson qubits \cite{Astafiev-Nat-449-588,RodrigueqImbersArmour-SSETMicromaser,Andre-SingleQubitLasingStrongCoupling,Andre-FewQubitLasingcQED} and  semiconductor quantum dot systems \cite{ChildressSorensen-MesoscopicCQED,JinMarthalerGolubev-NoiseSpectrumQuantumDotLaser, Liu-DoubleDotLaser-Science347.285.2015}.
In these setups, the low number of atoms is compensated by strong coupling.
Their frequencies are in the GHz regime, accordingly they
are sometimes called ``maser'' instead of laser. 
These systems may find useful applications, e.g., as miniaturized on-chip sources of coherent microwaves
in low temperature experiments. 
There are other on-chip microwave sources which have been studied, including
voltage-biased Josephson junctions \cite{Hoffheinz_Exp,Ankerhold_First,Juha_two} or
nonlinear resonators close to the quantum regime \cite{French_Nonlinear,Devoret_one,MM1}. However these devices  emit incoherent radiation unless driven by a coherent microwave source. Lasing devices based on  qubits also show unconventional properties, such as dressed-state lasing \cite{HaussFedorov-SingleQubitLasing, OelsnerMacha-DressedStateFluxQubitLasing}
or photon-number squeezed light \cite{Marthaler_Squeezed,SqueezedPhotonDistribution}.

So far, experimental realizations of lasers based on superconducting or quantum dot qubits have only used single or few artificial atoms.
A way to reach higher output power is to increase the number $M$ of atoms, e.g. by using superconducting quantum metamaterials.
Such materials with  $10-100$ qubits have already been produced and studied \cite{Metamaterial_Pascal}.
A drawback of using solid-state circuits is the fact that they invariably suffer from disorder, either due to the variations in the fabrication process or in the environment. As a result, the level-splitting $\epsilon_j$ 
($j=1,\dots M$)  and hence the detuning from the resonator frequency $\Delta_j = \epsilon_j/\hbar - \omega$,
as well as other parameters such as the coupling strength to the resonator or the local driving, vary for 
different artificial atoms. 
We therefore examine the influence of this (quasi-static) disorder on the multi-atom lasing.

Our analysis reveals that the lasing process is rather robust against disorder over a wide range of disorder variances. The origin of this effect is the following: Those atoms which are above the lasing threshold, e.g., close enough to resonance, start emitting photons into the resonator. These photons enhance the process of stimulated emission, which is proportional to their number $\erw{n}$, also for atoms which are still below the threshold. Hence these atoms start participating in the lasing process, and $\erw{n}$ grows further. In parallel to this growth, also the range of parameters such as detuning or coupling strength which are sufficient for lasing increases. 

In this paper, we will study the effects of disorder in the detuning, the coupling strength, and the pumping strength of the individual atoms on the photon number and the lasing thresholds. After presenting the model and the basic relations, we first compare the single- and the multi-atom setups of an ordered system. At this stage we observe already the increase of the allowed range of detuning and reduced requirement on the coupling strength. In the next sections quantitative results are presented for a Gaussian and a box distribution of the disorder in the various parameters. Since probably most experiments in the near future will be carried out with not too large numbers of atoms ($M\lesssim 100$) one should expect significant sample-specific deviations from the average behavior. We therefore also study these statistical properties. In Sect.~\ref{sec:SelfConsistentAddition} we reformulate the problem, which provides further insight into the mechanism responsible for the enhanced stability against disorder. 

\section{The Model}
We start from the Tavis-Cummings Hamiltonian for $M$ two-level systems (atoms), labeled by $j=1, \dots M$, with  level-splitting energy $\epsilon_j$ and coupled with strength $g_j$ to a common resonator with frequency $\omega$,
\begin{align}
	H = \hbar \omega a^\dagger a + \sum_{j=1}^M \frac{1}{2} \epsilon_j \tau_z^j + \sum_{j=1}^M \hbar g_j \left( \tau_+^j a + \tau_-^j a^\dagger \right)~.
\end{align}
Here $a$ and $a^\dagger$ are the photon operators of the radiation field and $\tau_{\nu}^j$
are Pauli matrices acting on the two states of atom $j$.

The influence of a dissipative  environment, leading to relaxation and decoherence processes is conveniently accounted for by a quantum master equation for the density matrix $\rho$ with the appropriate Lindblad terms
\begin{align}
	\frac{\d}{\d t} \rho &= - \frac{i}{\hbar} \left[ H, \rho \right] + L_\mathrm{R} \rho + \sum_{j=1}^M L_{\mathrm{Q},j} \rho ~, \\
	L_\mathrm{R} \rho 
	&= \frac{\kappa}{2} \left( 2 a \rho a^\dagger - a^\dagger a \rho - \rho a^\dagger a \right) ~, \\
	L_{\mathrm{Q},j} \rho 
	&= \frac{\Gamma_{\downarrow,j}}{2} \left( 2 \tau_-^j \rho \tau_+^j - \rho \tau_+^j \tau_-^j - \tau_+^j \tau_-^j \rho \right)   \\
	&+ \frac{\Gamma_{\uparrow,j}}{2} \left( 2 \tau_+^j \rho \tau_-^j - \rho \tau_-^j \tau_+^j - \tau_-^j \tau_+^j \rho \right) \nonumber \\ \nonumber
	&+ \frac{\Gamma_{\varphi,j}^*}{2} \left( \tau_z^j \rho \tau_z^j - \rho \right)~.
\end{align}
Here $\Gamma_{\uparrow,j}$ and $\Gamma_{\downarrow,j}$ are the effective excitation and decay rates of atom $j$, $\Gamma_{\varphi,j}^*$ its pure dephasing rate, and $\kappa$ the decay rate of the resonator.
Their combination defines the relaxation rate $\Gamma_{1,j} = \Gamma_{\uparrow,j} + \Gamma_{\downarrow,j}$ 
and the total dephasing rate $\Gamma_{\varphi,j} = \Gamma_{1,j}/2 + \Gamma_{\varphi,j}^*$. 
Below, we use the abbreviation $\Gamma_{\kappa,j} = \Gamma_{\varphi,j} + \kappa/2$. 
By adjusting the rates $\Gamma_{\uparrow,j} > \Gamma_{\downarrow,j}$ we can also account for the pumping of the lasing medium. Its strength is characterized by the value of the atomic polarization in equilibrium, $D_0 = (\Gamma_{\uparrow,j} - \Gamma_{\downarrow,j})/\Gamma_{1,j}$. 

From the master equation we obtain the generalized Maxwell-Bloch equations by using what is known as semi-quantum approximation \cite{Mandel-FluctuationLaserTheories}. It amounts to neglecting direct interatomic couplings , $\erw{\tau_+^i \tau_-^j} \approx 0$ for $i \ne j$, and factorizing $\erw{\tau_z^i a^\dagger a} \approx \erw{\tau_z^i} \erw{a^\dagger a}$. This yields
\begin{align}
	\frac{\d}{\d t} \erw{a^\dagger a} &= \sum_{j=1}^M i g_j \left( \erw{\tau_+^j a} - \erw{\tau_-^j a^\dagger} \right) - \kappa \erw{a^\dagger a} ~, 
	\label{eqn:EOMexpValN} \\
	\frac{\d}{\d t} \erw{\tau_z^j} &= 2 i g_j \left( \erw{\tau_-^j a^\dagger} - \erw{\tau_+^j a} \right) 
	\label{eqn:EOMexpValTauZ} \\ \nonumber
	&\quad - \Gamma_{1,j} \left( \erw{\tau_z^j} - D_{0,j} \right)~, \\
	\frac{\d}{\d t} \erw{\tau_+^j a} &= - \left( \Gamma_{\varphi,j} + \frac{\kappa}{2} - i \Delta_j \right) \erw{\tau_+^j a} 
	\label{eqn:EOMexpValTauPlusA} \\ \nonumber
	&\quad - i \frac{g_j}{2} \left( 1 + \left( 2 \erw{n} + 1 \right) \erw{\tau_z^j} \right)~.
\end{align}

In the stationary state, this set of equations can be cast into a fixed point equation for the quantum statistical  average photon number $\erw{n} = \erw{a^\dagger a}$,
\begin{align}
	\erw{n} = \sum_{j=1}^M \beta_j \frac{D_{0,j} \left( \erw{n} + \frac{1}{2} \right) + \frac{1}{2}}{\Gamma_{\kappa,j}^2 + \Delta_j^2 + \alpha_j \left( \erw{n} + \frac{1}{2} \right)}~.
	\label{eqn:FixpointEquationNGeneralCase}
\end{align}
Here we introduced the parameters
\begin{align*}
	\alpha_j = 4 g_j^2 \frac{\Gamma_{\kappa,j}}{\Gamma_{1,j}}~, \quad 
	\mbox{and} \quad
	\beta_j = 2 g_j^2 \frac{\Gamma_{\kappa,j}}{\kappa}~.
\end{align*}
The term $\erw{n}$  on the right-hand side of Eq.~\eqref{eqn:FixpointEquationNGeneralCase} accounts for the stimulated emission. 

\section{Photon number and allowed detuning of the ordered system}
\label{sec:PhotonNumberExpVal}

It is instructive to first consider the case without disorder. In this case, the fixed point equation has the solution
\begin{align}
	\erw{n}_M^0 &= X + \sqrt{X^2 + \frac{M \Gamma_1}{4 \kappa} (D_0 + 1)} ~, 
	\label{eqn:ExpValueNforMQubitsNoDisorder} \\
	X &= - \frac{1}{4} + \frac{M \Gamma_1}{4 \kappa} D_0 - \frac{\tilde{n}_0(\Delta)}{2} ~, \nonumber \\
	\tilde{n}_0(\Delta) &= \frac{\Gamma_1}{4 g^2} \frac{\Gamma_\kappa^2 + \Delta^2}{\Gamma_\kappa}~, \label{n0}
\end{align}
where the superscript $0$ refers to the absence of disorder. The quantity $\tilde{n}_0(\Delta)$ is the photon saturation number known already from the semi-classical theory of lasing 
\cite{Andre-SingleQubitLasingStrongCoupling,semiclassical}.
Within this theory the threshold for lasing is
\begin{align}
	\tilde{n}_0(\Delta) < \frac{M \Gamma_1}{2 \kappa} D_0 ~.
	\label{eqn:LasingThresholdConditionSCA}
\end{align}

\begin{figure}
	\centering
	\subfigure[]{
		\includegraphics[scale=.9]{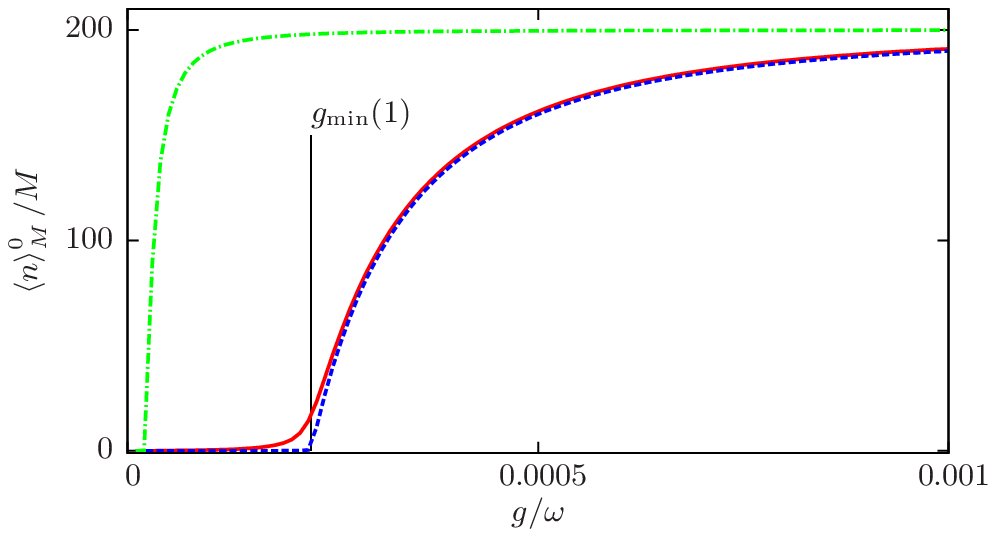}
		\label{fig:NoDisorderGErwN}
	}
	\subfigure[]{
		\includegraphics[scale=.9]{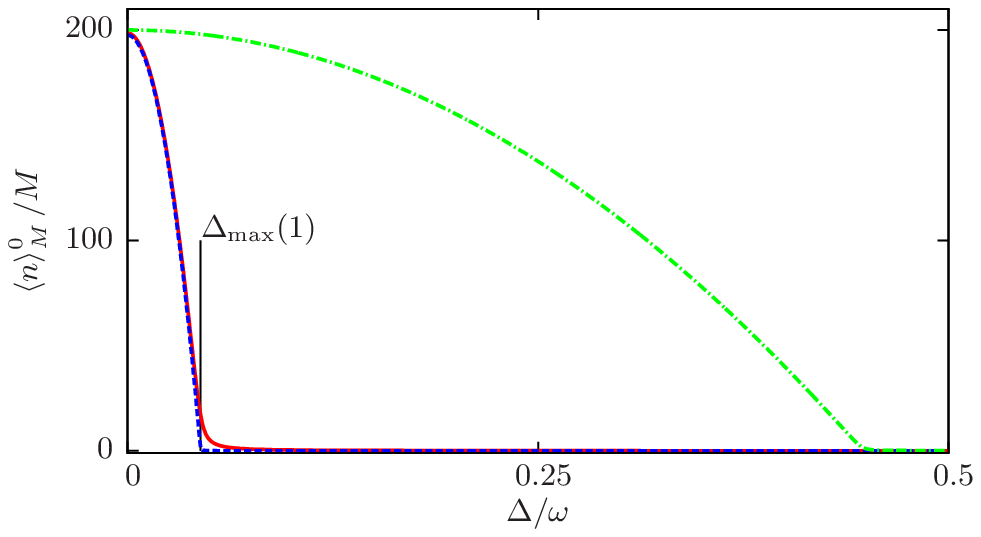}
		\label{fig:NoDisorderDeltaErwN}
	}
	\caption{(Color online) Properties of the stationary quantum statistical average photon number $\erw{n}_M^0(g,\Delta)/M$ per atom in an ordered lasing setup with $M$ atoms. \subref{fig:NoDisorderGErwN}: Plots as function of the coupling strength $g$ for atoms on resonance, $\Delta = 0$. \subref{fig:NoDisorderDeltaErwN}: Plots as function of the atomic detuning $\Delta$ at $g = 0.002$. 	
	Solid red curves represent $\erw{n}_1^0(g,\Delta)$, dashed blue curves $\erw{n}_M^0(\frac{g}{\sqrt{M}},\Delta)/M$ and dash-dotted green curves $\erw{n}_M^0(g,\Delta)/M$. Parameters are $\Gamma_\uparrow = 0.006$, $\Gamma_\downarrow = 0.002$, $\Gamma_\varphi^* = 0.001$, $\kappa = 0.00001$ and $M= 100$. All rates and couplings are measured in units of $\omega$.}
	\label{fig:NoDisorderTotalFigure}
\end{figure}

In Fig.~\ref{fig:NoDisorderTotalFigure} we display how the photon number $\erw{n}_M^0(g, \Delta)$ depends on the coupling strength $g$ and  detuning $\Delta$ 
for a many-atom setup with $M=100$ and the single-atom laser with $M=1$. 
Fig.~\ref{fig:NoDisorderTotalFigure}(a) illustrates the dependence 
on the coupling strength $g$. 
While the many-atom system shows a sharp, kink-like transition to the lasing state above $g_\mathrm{min}$, the transition is washed out for the single- and few-atom setup. However, even for $M=1$ it remains remarkably sharp. The crossover occurs at 
\begin{align}
	g_\mathrm{min}(M) &= \sqrt{\frac{1}{M} \frac{\kappa}{2 D_0} \frac{\Gamma_\varphi^2 + \Delta^2}{\Gamma_\varphi}} \propto \frac{1}{\sqrt{M}}.
	\label{gmin}
\end{align}
As a function of the detuning, illustrated in Fig.~\ref{fig:NoDisorderTotalFigure}(b), we note a gradual decrease of $\erw{n}_M^0(g, \Delta)$ with increasing $\Delta$, followed by a a sharp transition to a low, close to thermal
population. The crossover occurs at
\begin{align}
	\Delta_\mathrm{max}(M) &= \Gamma_\varphi \sqrt{2 g^2 M \frac{D_0}{\kappa \Gamma_\varphi} - 1} \propto \sqrt{M}~.
	\label{Deltamax}
\end{align}

The analytic results~\eqref{gmin} and~\eqref{Deltamax} follow from Eq.~\eqref{eqn:LasingThresholdConditionSCA}, which is, like Eq.~\eqref{n0}, obtained within the semi-classical approximation. Strictly, these approximations 
are valid for very large numbers of $M$, only. Remarkably they provide good estimates for the thresholds also for small $M$ \cite{Andre-FewQubitLasingcQED}.

The results shown in Fig.~\ref{fig:NoDisorderTotalFigure} also illustrate a remarkable scaling relation \cite{Andre-FewQubitLasingcQED}, namely
\begin{align}
	\frac{1}{M} \erw{n}_M^0 \left(\frac{g}{\sqrt{M}},\Delta\right) \approx \erw{n}_1^0(g,\Delta)~.
	\label{eqn:ScalingLawAndre}
\end{align}
The scaling relation, combined with Eqs.~\eqref{gmin} and \eqref{Deltamax}, displays the following 
properties: (i) in the lasing state the number of photons $\erw{n}_M^0$ grows linearly with the number of atoms $M$, (ii) for different $M$ we obtain the same qualitative dependence on the coupling strength provided that it is rescaled as $g/\sqrt{M}$, (iii) we furthermore note that 
the width of the resonance as function of detuning increases with the number of atoms proportional to $\sqrt{M}$. In other words, for a fixed coupling strength, a system with many atoms can tolerate a stronger detuning and still show a transition to the lasing state than a single-atom laser.  As we will show in the following section, this is the reason for the robustness against disorder which we observe. 

In the rest of the paper we focus on disordered systems. Their stationary quantum average photon number will be denoted by $\erw{n}_M$. 

\section{Disordered lasing medium}
\label{sec:DisorderAveragingProcedure}
We will now study the lasing transition in a system with many atoms $M \gg 1$ with disorder in either the coupling strength, the detuning or the pumping. Accordingly, we average Eq.~\eqref{eqn:FixpointEquationNGeneralCase} over the appropriate normalized probability distribution, e.g.,
\begin{align*}
	\sum_{i=1}^M \dots = M \iiint \d \Delta \,\d g\, \d D_0\, 
p(\Delta,g, D_0
) \dots ~.
\end{align*}
Having carried out the integration, we are left with a fixed point equation for  $\overline{\erw{n}}_M$, depending on $M$ and the distribution $p$. For clarity we concentrate in the following on disorder in only one lasing parameter at a time.

In general, the problem needs to be solved numerically. However, we can proceed analytically by using a Gaussian distribution $p_\mathrm{G}$ or a box distribution $p_\mathrm{B}$,
\begin{align*}
	p_\mathrm{G}(x) &= \frac{1}{\sqrt{2 \pi} \sigma_x} \exp \left( - \frac{1}{2} \frac{(x-\mu)^2}{\sigma_x^2} \right) ~, \\
	p_\mathrm{B}(x) &= \frac{1}{b} \left[ \Theta \left( x - \mu + \frac{b}{2} \right) - \Theta \left( x - \mu - \frac{b}{2} \right) \right]~, \\
	b &= \sqrt{12} \, \sigma_x ~,
\end{align*}
where $x$ is the variable to be averaged over, $\mu$ is its mean value and $\sigma_x$ its standard deviation. 

Below, we will present results for the disorder averages of the quantum statistical expectation values $\overline{\erw{n}}_M$.
On the other hand, an experimental realization of quantum metamaterial-based lasing most likely will not have a very large number $M$ of artificial atoms, and the probability distribution $p$ might not be sampled sufficiently to be well described by the integrals. Instead, for a given realization there will be deviations from the mean value $\overline{\erw{n}}_M$. We will analyze these fluctuations numerically by randomly generating setups of $M$ atoms with lasing parameters distributed according to a given probability distribution $p$, and solving Eq.~\eqref{eqn:FixpointEquationNGeneralCase} numerically for each of these setups. The solution for such a random system of $M$ atoms will be denoted by $\erw{n}_M$. They show variations from sample to sample.

\subsection{Disorder in the detuning}
In this subsection we examine disorder in the atomic detuning $\Delta = \epsilon/\hbar - \omega$. Its average is chosen to be zero. Since  $\epsilon \geq 0$ we have $\Delta \geq - \omega$. This constraints the width of the box distribution to $\sigma_\Delta \leq \omega/\sqrt{3}$. 
Also in the case of a Gaussian distribution 
we choose sufficiently narrow distributions to minimize the effect of unphysical values of the detuning. With these restrictions, we  average Eq.~\eqref{eqn:FixpointEquationNGeneralCase} analytically and obtain
\begin{align}
	\erw{n} &= M \beta \left[ D_0 \left( \erw{n} + \frac{1}{2} \right) + \frac{1}{2} \right] I \left( \zeta_{\erw{n}} \right)~, \\ \nonumber
	\zeta_{\erw{n}} &= \sqrt{\Gamma_\kappa^2 + \alpha \left( \erw{n} + \frac{1}{2} \right)}~.
\end{align}
The integrals $I(\zeta)$ for the two types of distributions are given by
\begin{align*}
	I_\mathrm{G}(\zeta) &= \sqrt{\frac{\pi}{2}} \frac{1}{\zeta \sigma_\Delta} \exp \left( \frac{\zeta^2}{2 \sigma_\Delta^2} \right) \erfc \left( \frac{\zeta}{\sqrt{2} \sigma_\Delta} \right) ~, \\
	I_\mathrm{B}(\zeta) &= \frac{1}{\sqrt{3} \zeta \sigma_\Delta} \arctan \left( \frac{\sqrt{3} \sigma_\Delta}{\zeta} \right) ~.
\end{align*}
Here, $\erfc$ is the complementary error function. 

\begin{figure}
	\includegraphics[scale=.9]{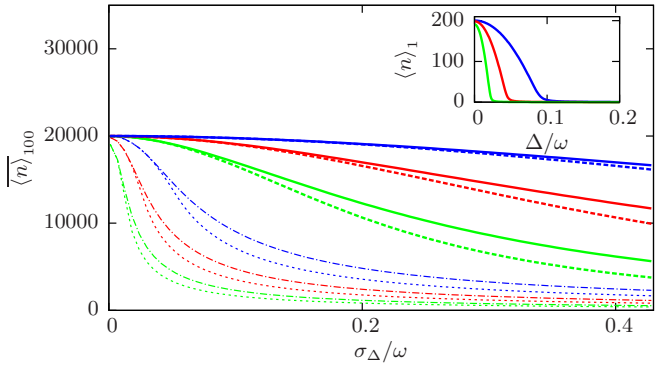}
	\caption{(Color online) Influence of disorder in the detuning $\Delta$ on the  quantum statistical average photon number $\overline{\erw{n}}_{100}$ in the resonator. Solid lines are calculated for a Gaussian distribution with mean $\overline{\Delta}=0$ and standard deviation $\sigma_\Delta$, dotted lines represent a box distribution with the same parameters. 
The average photon number decreases remarkably slowly with increasing disorder. This cannot be explained by averaging the single-atom resonance curves $\erw{n}_1(\Delta)$, examples of which are shown in the inset, over the same distribution of detuning. Averaging of the single-atom results would lead to the curves plotted with thin lines. 
Plot parameters are $M = 100$, $\Gamma_\uparrow = 0.006$, $\Gamma_\downarrow = 0.002$, $\Gamma_\varphi = 0.001$ and $\kappa = 0.00001$; blue curves $g=0.004$, red curves $g = 0.002$ and green curves $g = 0.001$. All rates and couplings are measured in units of $\omega$.}
	\label{fig:DisorderDeltaErwN}
\end{figure}

These stationary fixpoint equations can be solved numerically for $\overline{\erw{n}}_M(\sigma_\Delta)$. Results for $M=100$ atoms are shown by the thick upper curves in Fig.~\ref{fig:DisorderDeltaErwN}
for three values of $g$ and the two types of distribution $p_{G/B}(\Delta)$.  We note that disorder in the detuning decreases $\overline{\erw{n}}_{100}(\sigma_\Delta)$ only weakly over a broad range of the variance $\sigma_\Delta$. This behavior cannot be explained as an average of the single-atom laser resonances $\erw{n}_1(\Delta)$, which are displayed in the inset.
Averaging them over the distributions $p_{G/B}(\Delta)$ leads to the thin lower curves in Fig.~\ref{fig:DisorderDeltaErwN}, which obviously are much narrower in the disorder variance than the collective behavior of the $M$ atoms. We conclude that for disorder in the detuning, the setup with $M$ atoms shows lasing in a much broader range of detunings than what we obtain from the single-atom results $\erw{n}_1$ averaged over the same probability distribution. We will further illustrate this behavior in Sec.~\ref{sec:SelfConsistentAddition}.

In addition we observe that for weak coupling strength $g = 0.001\,\omega$, the average contribution of each atom in the $M$-atom setup in the limit $\sigma_\Delta \to 0$ is $\overline{\erw{n}}_{100}/100 \approx 200$, whereas $\erw{n}_1(\Delta=0) \approx 195$. That implies that the lasing activity per individual atom is enhanced in the $M$-atom setup as compared to the single-atom case, although for the parameters considered this is a rather weak effect. 

We also note that the results shown for the Gaussian and the box distribution nearly coincide. Both distributions were chosen to have the same average and second-order moment. In addition
 $\overline{\erw{n}}_M(\sigma_\Delta)$ depends only on even moments of $p_\mathrm{G/B}(\Delta)$. As a result, for sufficiently narrow distributions,  $\sigma_\Delta \ll \omega$, both distributions yield similar results.

\begin{figure}
	\includegraphics[scale=1]{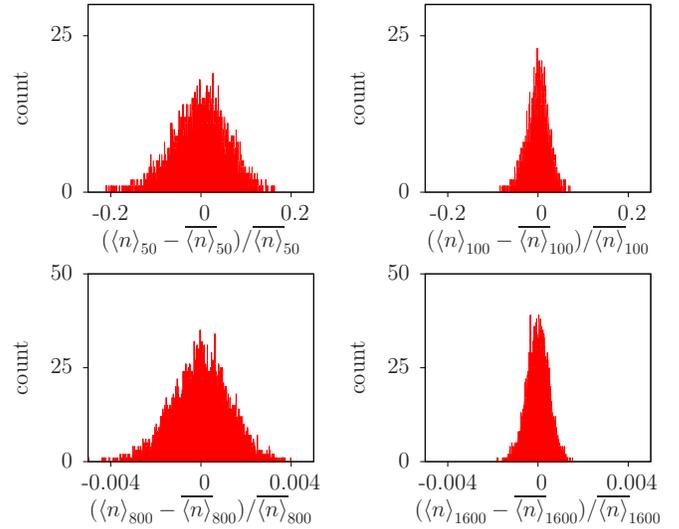}
	\caption{(Color online) Fluctuations $\erw{n}_M$ around $\overline{\erw{n}}_M$ for disorder in the detuning due to non-perfect sampling of a Gaussian disorder distribution with $\sigma_\Delta = 0.2$. Histograms are created for $10000$  systems, randomly chosen with the Gaussian distribution, for $M=50$, $100$, $800$, $1600$ atoms, respectively. Results are $\overline{\erw{n}}_{50} = 7479$, $\overline{\erw{n}}_{100} = 16976$, $\overline{\erw{n}}_{800} = 156184$, 
$\overline{\erw{n}}_{1600} = 316100$ with standard deviations $386$, $352$, $182$ and $136$ photons, respectively. Plot parameters are $\overline{\Delta} = 0$, $g = 0.002$, $\Gamma_\uparrow = 0.006$, $\Gamma_\downarrow = 0.002$, $\Gamma_\varphi^* = 0.001$, $\kappa = 0.00001$. All rates and couplings are measured in units of $\omega$.}
	\label{fig:DisorderDeltaHistogramme}
\end{figure}

We conclude this subsection with an analysis of the sample-to-sample variations of setups with a finite number $M$ of atoms. For this purpose, we consider ensembles with random parameters chosen according to the Gaussian distribution  $p_{G}(\Delta)$ and solve Eq.~\eqref{eqn:FixpointEquationNGeneralCase} numerically. Results of $\erw{n}_M$, varying around its mean value $\overline{\erw{n}}_M$,  are shown in Fig.~\ref{fig:DisorderDeltaHistogramme} for the variance $\sigma_\Delta = 0.2\,\omega$  and $M=50$, $100$, $800$ and $1600$ atoms.  On long enough time scales, when the quasi-static parameters vary in an experiment, we expect that the lasing intensity will vary accordingly.

\subsection{Disorder in the coupling strength}
Similar to the previous case, we examine disorder in the coupling strength $g$ between the atoms and the resonator. The condition $g \geq 0$ imposes rigorous constraints on the box distribution, $\sigma_g \leq \overline{g}/\sqrt{3}$, 
and similar approximate conditions for the 
Gaussian distribution.

After averaging, the fixed point equation~\eqref{eqn:FixpointEquationNGeneralCase} becomes 
\begin{align}
	\erw{n} &= \frac{M \Gamma_1}{2 \kappa} \left( D_0 + \frac{1}{2 \erw{n} + 1} \right) I \left( c_{\erw{n}} \right) ~,\\ \nonumber
	c_{\erw{n}} &= \sqrt{\frac{\Gamma_1(\Gamma_\kappa^2 + \Delta^2)}{4 (\erw{n} + 1/2) \Gamma_\kappa}} ~,
\end{align}
with the integrals
\begin{align*}	
	I_\mathrm{G}(c) &= 1 - \pi c  V(\overline{g},\sigma_g, c)~, \\
	I_\mathrm{B}(c) &= 1 - \frac{c}{b} \\
	&\quad \times \left[ \arctan \left( \frac{\mu + b/2}{c} \right) - \arctan \left( \frac{\mu - b/2}{c}\right)
 \right] ~.
\end{align*}
Again, we have $b = \sqrt{12} \, \sigma_g$, and $V$ is the Voigt function
\begin{align*}
	&V(\overline{g},\sigma_g,x) \\
	&= \sqrt{\frac{\pi}{2}} \frac{1}{2 \pi \sigma_g}  \exp \left[ \left( \frac{x - i \overline{g}}{\sqrt{2} \sigma_g} \right)^2 \right] \erfc \left( \frac{x - i \overline{g}}{\sqrt{2} \sigma_g} \right) + \mathrm{c.c.} ~.
\end{align*}

\begin{figure}
	\includegraphics[scale=.9]{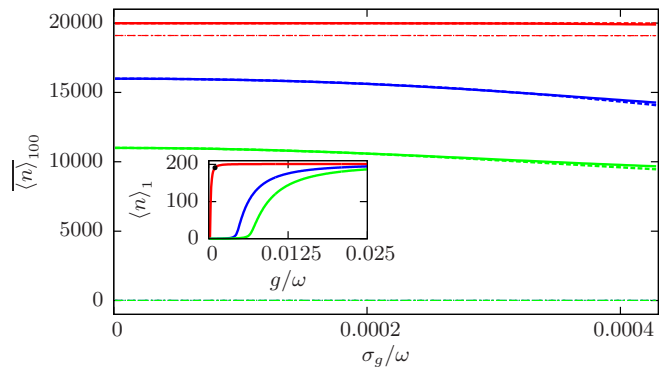}
	\caption{(Color online) Influence of disorder in the coupling strength $g$ on the stationary quantum statistical average photon number $\overline{\erw{n}}_{100}$ in the resonator (thick lines). 
	Solid lines are calculated for a Gaussian distribution with mean $\overline{g}=0.001\,\omega$ and standard deviation $\sigma_g$, dotted lines represent a box distribution with the same parameters. 
	The results cannot be explained by averaging the single-atom resonance curves $\erw{n}_1$ 
over the same distribution of coupling strength. This would yield the results represented by thin lines. The inset shows the corresponding single-atom lasing resonance curves $\erw{n}_1$. The average $\overline{g}$ is indicated by a black circle.  
	Plot parameters are $M = 100$, $\Gamma_\uparrow = 0.006$, $\Gamma_\downarrow = 0.002$, $\Gamma_\varphi = 0.001$ and $\kappa = 0.00001$. Red curves $\Delta=0$, blue curves $\Delta=0.1$ and green curves $\Delta=0.15$. The thin blue and green curves coincide.
	All rates and couplings are measured in units of $\omega$.}
	\label{fig:DisorderGErwN}
\end{figure}

Fig.~\ref{fig:DisorderGErwN} shows numerical solutions $\overline{\erw{n}}_M(\sigma_g)$ of these fixpoint equations for $M=100$ atoms at resonance, $\Delta = 0$, and for two non-zero values of atomic detuning (thick curves). The mean coupling strength $\overline{g} = 0.001\,\omega$ is chosen to be close to the lasing threshold of a single atom in resonance. The position of $\overline{g}$ is indicated at the single-atom resonance curve in the inset by a black dot.

For disorder in the coupling strength, similar to what we found above, the properties of $\overline{\erw{n}}_M$ cannot be explained by averaging the single-atom resonance curves $\erw{n}_1(g)$ over the distribution of $p_{G/B}(g)$ (which would result in the thin lines). The averaging yields a smaller average photon number because a part of the atoms are coupled with $g < \overline{g}$ and therefore do no contribute (significantly) to the lasing process. 
On the other hand, $\overline{\erw{n}}_{100}/100 \approx 200$, hence all atoms are actually participating in the lasing process at their maximum contribution irrespective of their actual individual coupling strength $g$. This effect is even more pronounced for $\Delta = 0.1$ or $\Delta = 0.15$ when the single-atom resonance curves and their na\"ive average predict no lasing activity at all. However, the multi-atom setup is still operating at approximately $80\,\%$ or $50\,\%$ of its maximum photon number, respectively. In Sec.~\ref{sec:SelfConsistentAddition} we will provide further explanations of these properties. 

In a typical lasing experiment, the pumping rates are chosen such that the laser operates far above the lasing threshold of a single resonant atom. Then, the decrease in $\overline{\erw{n}}_M$ for increasing $\Delta$ is even less pronounced than shown in Fig.~\ref{fig:DisorderGErwN}. In this figure we also observe again that the results obtained for a Gaussian and a box distribution coincide for a wide range of variance $\sigma_g$. 

\begin{figure}
	\includegraphics[scale=.95]{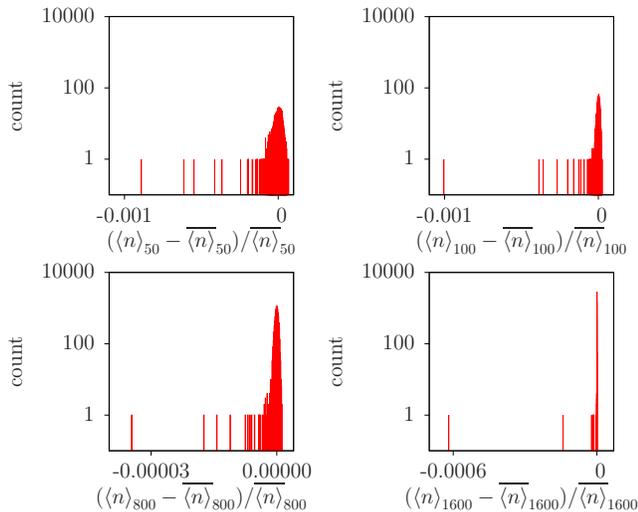}
	\caption{(Color online) Fluctuations $\erw{n}_M$ around $\overline{\erw{n}}_M$ for disorder in the coupling strength due to a non-perfect sampling of a Gaussian distribution with $\sigma_g = 0.0004\,\omega$. Histograms are created for $10000$ systems, randomly chosen with the Gaussian distribution, with $M = 50$, $100$, $800$, $1600$, respectively. $\overline{\erw{n}}_{50} = 9998$, $\overline{\erw{n}}_{100} = 19998$, $\overline{\erw{n}}_{800} = 159998$, $\overline{\erw{n}}_{1600} = 319998$. Plot parameters are $\overline{g} = 0.002$, $\Gamma_\uparrow = 0.006$, $\Gamma_\downarrow = 0.002$, $\Gamma_\varphi^* = 0.001$, $\kappa = 0.00001$. All rates and couplings are measured in units of $\omega$.}
	\label{fig:DisorderGHistogramme}
\end{figure}

We conclude with an analysis of the fluctuations in $\erw{n}_M$ around the mean value $\overline{\erw{n}}_{M}$ due to disorder in $g$ for a finite system size $M$, by the same procedure as described above for the variations in the detuning. The histograms of $\erw{n}_M$ in Fig.~\ref{fig:DisorderGHistogramme} show a main peak around $\overline{\erw{n}}_M$ and a tail representing a few ensembles with much lower $\erw{n}_M$. This tail arises because we choose $g$ close to the lasing transition: Some atoms have such weak coupling strengths that they cannot participate in the lasing process. The corresponding systems have effectively a reduced $M$. As these systems occur rarely, an ensemble of $10000$ systems is not sufficient to produce a smooth distribution. The standard deviation of the main peak is $0.24$, $0.17$, $0.07$ and $0.05$ photons, respectively.

\subsection{Disorder in the pumping}
The stationary value of the atomic polarization $D_0 = (\Gamma_\uparrow - \Gamma_\downarrow)/(\Gamma_\uparrow + \Gamma_\downarrow)$ is a function of the pumping and relaxation rates. These rates appear in Eq.~\eqref{eqn:FixpointEquationNGeneralCase} via the expressions $D_0$ but also $\Gamma_1 = \Gamma_\uparrow + \Gamma_\downarrow$. In this subsection, we concentrate on the effect of disorder in $D_0$ while assuming that $\Gamma_1$ is fixed. A motivation is provided by the system analyzed in Ref.~\onlinecite{Astafiev-Nat-449-588}, where
\begin{align*}
	\Gamma_\uparrow - \Gamma_\downarrow &\propto \cos(\theta)~, \\
	\Gamma_1 = \Gamma_\uparrow + \Gamma_\downarrow &\propto \frac{1}{2} \left( 1 + \cos^2(\theta) \right)~,
\end{align*}
and $\theta$ is the mixing angle of Coulomb and Josephson energy. Typically, $\theta \lesssim \pi/2$, so that $D_0$ fluctuates proportional to $\theta-\pi/2$ whereas $\Gamma_1$ is approximately constant.

Averaging Eq.~\eqref{eqn:FixpointEquationNGeneralCase} over the disorder in $D_0$, we arrive at the fixpoint equation  
\begin{align}
	\erw{n} = &\frac{M \beta \left[\overline{D}_0 \left( \erw{n} + \frac{1}{2} \right) + \frac{1}{2} \right]}{\Gamma_\kappa^2 + \Delta^2 + \alpha (\erw{n} + \frac{1}{2})}~.
\end{align}
Its solution is of the same form as Eq.~\eqref{eqn:ExpValueNforMQubitsNoDisorder}, except that $D_0$ is replaced by $\overline{D}_0$. 
For typical lasing parameters above the threshold, Eq.~\eqref{eqn:LasingThresholdConditionSCA} and 
the relation $\Gamma_1 \overline{D}_0 \gg \kappa$ hold. Then, Eq.~\eqref{eqn:ExpValueNforMQubitsNoDisorder} reduces to a linear dependence on $\overline{D}_0$, 
\begin{align*}
	\erw{n}_M = \frac{M \Gamma_1}{2 \kappa} \overline{D}_0 - \tilde{n}_0(\Delta)~.
\end{align*}
This means that, in contrast to the previously discussed examples, disorder in $D_0$ is properly accounted for by averaging  over the single-atom results  $\erw{n}_1$. 

\section{Discussion}
\label{sec:SelfConsistentAddition}
The physical origin for the robustness of the system against disorder is an increased stimulated emission of each individual atom if there are additional photons $\erw{n_\mathrm{add}}$ in the cavity originating from the lasing activity of other atoms. 
For ordered systems, the enhanced lasing activity, i.e., the growth of the average photon number, $\erw{n}_M \propto M$, and the increased range of allowed detuning is explicitly derived from Eq.~\eqref{eqn:FixpointEquationNGeneralCase}. For disordered systems with disorder in the detuning or coupling strength, we found similar behavior of the average quantity $\overline{\erw{n}}_M$. 
To gain further insight how $\erw{n_\mathrm{add}}$ additional photons in the resonator broaden and enhance the lasing activity of each individual atom $\erw{n_i}$, we split 
\begin{align}
	\erw{n} = \sum_{j=1}^M \erw{n_j} = \erw{n_i} + \erw{n_\mathrm{add}^i} 
	\label{eqn:SelfConsistentDefTotalPhotonNumber}
\end{align}
with
\begin{align}
	\erw{n_\mathrm{add}^i} = \sum_{\stackrel{j = 1}{j \neq i}}^M \erw{n_j}~.
	\label{eqn:SelfConsistentDefAddPhotonNumber}
\end{align}
Accordingly, we split Eq.~\eqref{eqn:EOMexpValN} into $M$ equations for $\erw{n_i}$, what can be interpreted as the contributions of atom, $i$, which is solved by
\begin{align}
	\erw{n_i} = \beta_i \frac{D_{0,i} (\erw{n_i} + \erw{n_\mathrm{add}^i} + 1/2) + 1/2}{\Gamma_{\kappa,i}^2 + \Delta_i^2 + \alpha_i (\erw{n_i} + \erw{n_\mathrm{add}^i} + 1/2)} \, ,
	\label{eqn:split}
\end{align}
If we sum this relation over all $i=1, \dots M$ we recover Eq.~\eqref{eqn:FixpointEquationNGeneralCase}, but it is also valid for the single-atom case,  $M=1$. 
Since on the right hand side $\erw{n_\mathrm{add}^i}$ appears always together with $\erw{n_i}$, which is the term describing for a single atom the stimulated emission, the relation displays the property that each individual atom acquires an enhanced lasing activity by the presence of additional photons in the resonator. 

Eq.~\eqref{eqn:split} is solved by
\begin{align}
	\erw{n_i} &= \tilde{X} + \sqrt{\tilde{X}^2 + \frac{\Gamma_{1,i}}{2 \kappa} \nonumber \left[D_{0,i} \left( \erw{n_\mathrm{add}^i} + \frac{1}{2} \right) + \frac{1}{2} \right]}~, \\
	\tilde{X} &= - \frac{1}{4} - \frac{\erw{n_\mathrm{add}^i}}{2} + \frac{\Gamma_{1,i}}{4 \kappa} D_{0,i} - \frac{\tilde{n}_0}{2}~. 
	\label{eqn:SelfConsistentContributionAtomI}
\end{align}
This set of $M$ equations for $\erw{n_i}$ has to be solved self-consistently via the relation Eq.~\eqref{eqn:SelfConsistentDefTotalPhotonNumber}. This is what we had done (effectively) in the previous parts of the paper both for an ordered as well as for disordered systems.

\begin{figure}
	\centering
	\subfigure[]{
		\includegraphics[scale=.9]{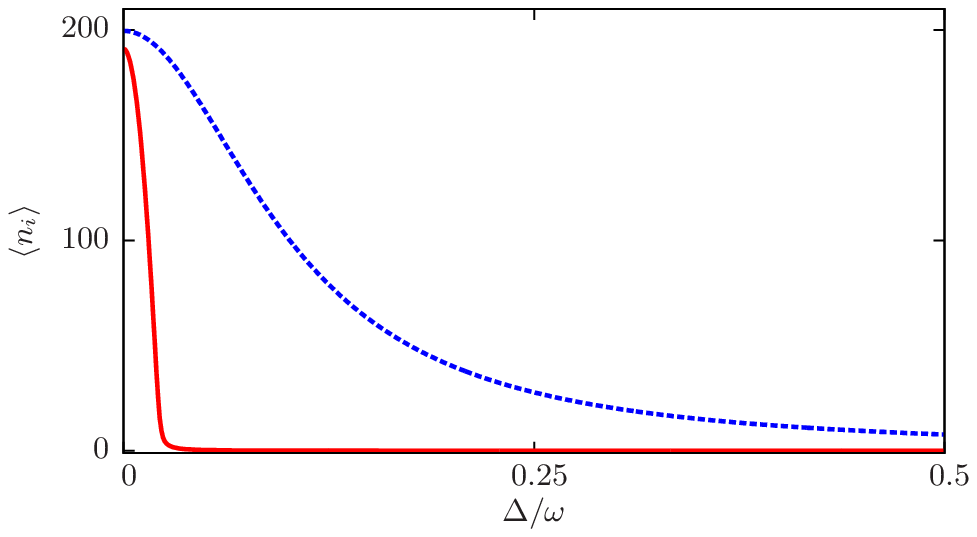}
		\label{fig:BroadenedResonanceCurveDelta}
	}
	\subfigure[]{
		\includegraphics[scale=.9]{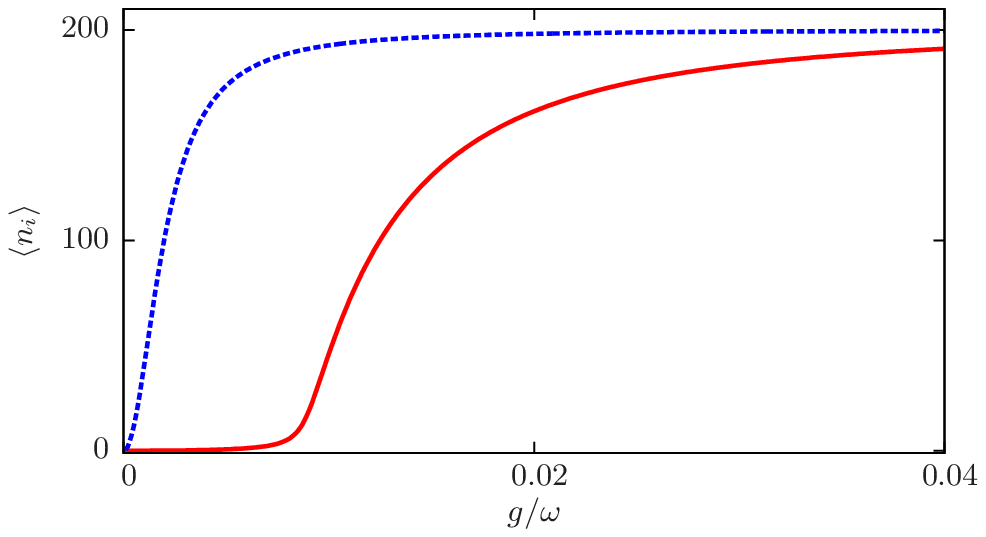}
		\label{fig:BroadenedResonanceCurveG}
	}
	\caption{(Color online) \subref{fig:BroadenedResonanceCurveDelta}: $\erw{n_i}$ as a function of detuning for fixed $g=0.001$. \subref{fig:BroadenedResonanceCurveG}: $\erw{n_i}$ as a function of coupling strength for fixed $\Delta = 0.2$. Solid red curves represent the case without additional photons in the resonator, dashed blue curves the same in the presence of $\erw{n^i_\mathrm{add}} = 4000$ additional photons in the resonator cavity. Their presence increases the lasing activity of each individual atom as well as the range of allowed detuning, and decreases the threshold in coupling strength. Plot parameters are $\Gamma_\uparrow=0.006$, $\Gamma_\downarrow=0.002$, $\Gamma_\varphi^*=0.001$, $\kappa=0.00001$.}
	\label{fig:BroadenedResonanceCurveGeneralFigure}
\end{figure}

To illustrate the enhancement effect we can also simply assume that there exist additional photons,
wherever they come from.
Fig.~\ref{fig:BroadenedResonanceCurveGeneralFigure} compares plots of $\erw{n_i}$ for $\erw{n_\mathrm{add}^i} = 0$ and $4000$. Fig.~\ref{fig:BroadenedResonanceCurveDelta}
demonstrates the broadening of the lasing resonance curve as a function of the detuning due to the presence of additional photons, as well as a slight enhancement at $\Delta = 0$, consistent with the observations made in Fig.~\ref{fig:DisorderDeltaErwN}. Fig.~\ref{fig:BroadenedResonanceCurveG}
demonstrates the 
lowering of the lasing threshold coupling strength, consistent with the observations made in Fig.~\ref{fig:DisorderGErwN}.

For disordered setups with a finite-width disorder distribution, we observed that atoms above the lasing threshold, e.g., close enough to resonance, ``drag'' others, which appeared to be below, also into a lasing state, and finally a self-organized stationary state is established. But we found an enhancement of the lasing window also in the case of ordered systems. Here we like to point out that the reformulation presented in this section can reproduce also this property. To understand it we consider  $M$ identical atoms which would all be off-resonant in the single-atom setup, $\Delta > \Delta_\mathrm{max}(1)$. It is important to note that the semi-quantum model does not exhibit a sharp transition to the lasing state. Therefore, each of the atoms produces a small but non-vanishing contribution $\erw{n_i}$ to the total photon number $\erw{n}_M^0$. This possibly very small contribution is then enhanced by the presence of all other ones, which may be sufficient to drive the system into the self-consistent broadened state.

\section{Conclusion}
In this paper, we showed that a multi-atom lasing setup with $M \gg 1$ atoms is rather robust against disorder in the individual atomic parameters, e.g.\ detuning, coupling strength to the resonator, or pumping strength. If an atom is coupled to a cavity that contains additional photons not originating from the atom itself, its lasing activity is nevertheless enhanced due to stimulated emission. This leads to a growth of the number of photons scaling with $M$ but also to a broadening of the resonance conditions, with the maximum allowed detuning scaling proportional to $\sqrt{M}$. Therefore, multiple atoms connected to a common resonator can effectively drag each other into resonance and generate a self-consistent stationary state that is robust against disorder. 

The average total photon number $\overline{\erw{n}}_M$ of the setup can be calculated by performing the averages 
implied by the fixed point equation~\eqref{eqn:FixpointEquationNGeneralCase}. We have performed these averages for two types of distributions, box and Gaussian, with similar results. Since currently systems with relatively low numbers of artificial atoms ($M \le 100 $) are of interest for lasing experiments we also examined the 
fluctuations around $\overline{\erw{n}}_M$ due to the imperfect sampling of the parameter distribution. This provides estimates for sample-to-sample fluctuations in such lasing setup, as well.

The conclusion from our analysis is that imperfections in the control of material parameters do not prohibit the construction of multi-atom lasing setups. 
This will help the construction of miniaturized on-chip radiation sources for low-temperature microwave experiments. 

\section{Acknowledgements}
We acknowledge fruitful discussions with J.\ Braum\"uller. This work was supported by the DFG Research Grants SCHO 287/7-1 and MA 6334/3-1.


\begin{thebibliography}{999}
\bibitem{LambScully-LaserPhysics}
W.\ E.\ Lamb, W.\ P.\ Schleich, M.\ O.\ Scully, and C.\ H.\ Townes, Rev.\ Mod.\ Phys.\ \textbf{71}, S263 (1999).

\bibitem{ScullyZubairy-QuantumOptics}
M.\ O.\ Scully, M.\ S.\ Zubairy, \emph{Quantum Optics}, Cambridge University Press, 1997.

\bibitem{ThompsonSemiconductorLaser}
G.\ H.\ B.\ Thompson, \emph{Physics of semiconductor laser devices}, Wiley, 1980.

\bibitem{Astafiev-Nat-449-588}
O.\ Astafiev, K.\ Inomata, A.\ O.\ Niskanen, T.\ Yamamoto, Yu.\ A.\ Pashkin, Y.\ Nakamura, and J.\ S.\ Tsai, Nature \textbf{449}, 588 (2007).

\bibitem{Andre-FewQubitLasingcQED}
S.\ Andr\'e, V.\ Brosco, A.\ Shnirman, and G.\ Sch\"on, Phys.\ Rev.\ A  \textbf{79}, 053848 (2009).

\bibitem{Andre-SingleQubitLasingStrongCoupling}
S.\ Andr\'e, P.-Q.\ Jin, V.\ Brosco,\ J.\ H.\ Cole, A.\ Romito, A.\ Shnirman, and G.\ Sch\"on, Phys.\ Rev.\ A \textbf{82}, 053802 (2010).

\bibitem{RodrigueqImbersArmour-SSETMicromaser}
D.\ A.\ Rodrigues, J.\ Imbers, and A.\ D.\ Armour, Phys.\ Rev.\ Lett.\ \textbf{98}, 067204 (2007).

\bibitem{ChildressSorensen-MesoscopicCQED}
L.\ Childress, A.\ S.\ S\o{}rensen, and M.\ D.\ Lukin, Phys.\ Rev.\ A \textbf{69}, 042302 (2004).

\bibitem{JinMarthalerGolubev-NoiseSpectrumQuantumDotLaser}
P.\ Q.\ Jin, M.\ Marthaler, J.\ H.\ Cole, A.\ Shnirman, and G.\ Sch\"on, Phys.\ Rev.\ B  \textbf{84}, 035322 (2011); J.\ Jin, M.\ Marthaler, P.\ Q.\ Jin, D.\ Golubev, and G.\ Sch\"on, New.\ J.\ Phys.\ \textbf{15}, 025044 (2013).

\bibitem{Liu-DoubleDotLaser-Science347.285.2015}
Y.-Y.\ Liu, J.\ Stehlik, C.\ Eichler, M.\ J.\ Gullans, J.\ M.\ Taylor, and J.\ R.\ Petta,
Science {\bf 347}, 285 (2015).

\bibitem{Hoffheinz_Exp}
M.\ Hofheinz, F.\ Portier, Q.\ Baudouin, P.\ Joyez, D.\ Vion, P.\ Bertet, P.\ Roche, and D.\ Esteve,
Phys.\ Rev.\ Lett.\ {\bf 106}, 217005 (2011).

\bibitem{Ankerhold_First}
V.\ Gramich, B.\ Kubala, S.\ Rohrer, and J.\ Ankerhold,
Phys.\ Rev.\ Lett.\ {\bf 111}, 247002 (2013).


\bibitem{Juha_two}
J.\ Lepp\"akangas, M.\ Fogelstr\"om, A.\ Grimm, M.\ Hofheinz, M.\ Marthaler, and G.\ Johansson,
Phys.\ Rev.\ Lett.\ {\bf 115}, 027004 (2015). 


\bibitem{French_Nonlinear}
F.\ Mallet, F.\ R.\ Ong, A.\ Palacios-Laloy, F.\ Nguyen, P.\ Bertet, D.\ Vion, and D.\ Esteve,
Nat.\ Phys.\ {\bf 5}, 791 (2009). 

\bibitem{Devoret_one}
A.\ Kamal, J.\ Clarke, and M.\ H.\ Devoret,
Nat.\ Phys.\ {\bf 7}, 311 (2011).


\bibitem{MM1}
S.\ Andre, L.\ Guo, V.\ Peano, M.\ Marthaler, and G.\ Sch\"on,
Phys.\ Rev.\ A {\bf 85}, 053825 (2012).



\bibitem{HaussFedorov-SingleQubitLasing}
J.\ Hauss, A.\ Fedorov, C.\ Hutter, A.\ Shnirman, and G.\ Sch\"on, Phys.\ Rev.\ Lett.\ \textbf{100}, 037003 (2008).

\bibitem{OelsnerMacha-DressedStateFluxQubitLasing}
G.\ Oelsner, P.\ Macha, O.\ V.\ Astafiev, E.\ Il'ichev, M.\ Grajcar, U.\ H\"ubner, B.\ I.\ Ivanov, P.\ Neilinger, and H.-G.\ Meyer,
Phys.\ Rev.\ Lett.\ \textbf{110}, 053602 (2013).

\bibitem{Marthaler_Squeezed}
M.\ Marthaler, J.\ Lepp\"akangas, and J.\ H.\ Cole, Phys.\ Rev.\ B {\bf 83}, 180505 (2011).

\bibitem{SqueezedPhotonDistribution}
M.\ Koppenh\"ofer and M.\ Marthaler, arXiv:1510.04168 (2015).

\bibitem{Metamaterial_Pascal}
P.\ Macha, G.\ Oelsner, J.-M.\ Reiner, M.\ Marthaler, S.\ Andr\'e, G.\ Sch\"on, U.\ H\"ubner, H.-G.\ Meyer, E.\ Il'ichev, and A.\ V.\ Ustinov,
Nat.\ Comm.\ {\bf 5}, 5146 (2014).

\bibitem{Mandel-FluctuationLaserTheories}
P.\ Mandel, Phys.\ Rev.\ A \textbf{21}, 2020 (1980).

\bibitem{semiclassical} The semi-classical approximation is valid in the limit of very large photon numbers,
but the parameter 
$\tilde{n}_0(\Delta)$ appears also in the present semi-quantum approximation used to study systems with finite $M$.

\end{thebibliography}
\end{document}